\documentclass{article}

\usepackage{PRIMEarxiv}

\usepackage[utf8]{inputenc} 
\usepackage[T1]{fontenc}    
\usepackage{hyperref}       
\usepackage{url}            
\usepackage{booktabs}       
\usepackage{amsfonts}       
\usepackage{nicefrac}       
\usepackage{microtype}      
\usepackage{lipsum}
\usepackage{fancyhdr}       
\usepackage{graphicx}       

\graphicspath{{media/}}     

\title{Unsupervised Learning Algorithms for Keyword Extraction in Undergraduate Thesis
}

\author{
  Fred Torres-Cruz* , Edelfré Flores, William E. Arcaya, Irenio L. Chagua, Marga I. Ingaluque \\
  Maestría en Ingeniería de Sistemas\\
  Universidad Nacional del Altiplano   \\
  Puno- Perú\\
  ftorres[at]unap.edu.pe \\
   \And
}

\begin{document}
\maketitle

\begin{abstract}
The amount of data managed in many academic institutions has increased in recent years, particularly in all the research work done by undergraduate students, who simply use empirical techniques for keyword selection, forgetting existing technical methods to assist their students in this process. Information and communication technologies, such as the platform for integrated research and academic work with responsibility (PILAR), which records information about research projects, such as titles, summaries, and keywords in their various modalities, have gained relevance and importance in the management of these. We proved algorithms with these records of research projects that have been analysed in this study, and predictions were made for each of the nine (09) models of unsupervised machine learning algorithms that were implemented for each of the 7430 records from the dataset. The most efficient way of extracting keywords for this dataset was the TF-IDF method, obtaining 72\% accuracy and [0.4786, SD 0.0501] in average extraction time for each thesis file processed by this model. 
\end{abstract}

\keywords{Keyword extraction\and Machine learning\and thesis \and university students \and unsupervised learning }

\section{Introduction}
It is undeniable the increase in the volumes of information that are being generated through the implementation of knowledge information systems in organizations, with the aim of achieving efficient management and service, on the other hand, universities are implementing services with the help of technology of different types that allow transactions at different levels of institutional management storing information in all its processes \cite{ranguelov2012a} ,\cite{mart2020a}.  All the existing volume of information on the Internet is growing permanently and acquires different forms of representation, from simple text files on a personal computer or an electronic newspaper to digital libraries and much larger and complex spaces such as the web, this information in hordes increased even more with the use of digital media  \cite{tolosa2008a}, \cite{jayawardene2021a} , the Universidad Nacional del Altiplano de Puno is no stranger to this affluent growth, so from the implementation of information systems such as the Platform for Integrated Research to Academic Work with Responsibility\cite{torres-cruz2016a} (PILAR), as well as the other like research platform for the special fund for university teachers, through these applications, information is being collected as well as the management of knowledge, generated by students, graduates, but noting in their procedures no treatment at the time of choosing the keywords.

Automatic keyword selection approaches are increasingly important to classify large volumes of documents, this process has become essential to make these documents more manageable and to obtain valuable information \cite{ahadh2021a}. Automatic keyword extraction helps to filter and find better recommendation and retrieval of information based on the content of the text itself, thus has a representation based on the content being evaluated \cite{ahadh2021a} .The goal of automatic keyword extraction is the application of the power and speed of current computational and computing capabilities to the problem of access and retrieval, with emphasis on information organization \cite{florescu2017a}. Likewise for information and knowledge extraction is the subject of considerable research interest in the fields of machine learning and data mining. text data mining and in particular text mining has become one of the most active sub fields of research in data mining \cite{xuezhong2010a}.

Therefore, the present work addresses the use of the most representative techniques of automatic keyword extraction, using unsupervised machine learning models, being these techniques one of those that are rapidly implemented, as well as laying the groundwork for other similar studies and that this area of knowledge can continue to be studied.

\section{Literature Review}
\subsection{Definitions}
\subsubsection{Unsupervised learning}

Unsupervised learning is the equivalent of grouping and this process is not subject to a contrast analysis, so this grouping is not previously identified, if we go to a specific section in the extraction of keywords with the unsupervised approach it is not necessary to need a manual or automatic labelling corpus.  To perform these tasks there are different techniques in which the characteristics of the text are analysed to obtain a better extraction effect \cite{xu2021a} . 

\subsubsection{Keyword selection}

For the selection of keywords for research papers, articles and academic documents, the author must choose between 3 to 10 words \cite{gonz2012a} the same ones that should represent the idea that most of the time are presented in the abstract and title of the work repeatedly,  for easy location and traceability\cite{mack2012a}. After this main ide of keyword selection, we may talk about.	 

\subsubsection{Keyword Extraction}

Depending on the model, autonomous keyword extraction is the process of selecting words and phrases from a text document that, at best we may project the central idea of the document without any human intervention \cite{zhang2008a}. 

\subsection{Background}
The extraction of words is widely studied from the different aspects of computer science in particular from machine learning, for which we present the following works that support the development of this research work like Xu \& Zhang who propose a new approach to extracting keywords from a text that combines characteristics such as word frequency and the association between them \cite{xu2021a}, in other field Kretschmann et al, develop a system for keyword assignment for scientific abstracts to be functional as recommendations of researchers\cite{kretschmann2020a}. YAKE is an unsupervised learning keyword extraction method that relies on statistical text characteristics extracted from individual documents \cite{campos2020a}. 

As we can see many authors has been work in this field of study such as Guan et al. who show us  in their research how improved TF-IDF for article keywords extraction specifically the accuracy of the algorithm\cite{guan2019a}, like them other representative study was made by Mahata et al., writing an article on keyword extraction called Key2Vec with unsupervised methods by taking advantage of the formation of multi-word phrase embeddings that are used for the thematic representation of scientific articles \cite{mahata2018a} , same as these examples that we show we may prove these methos an get an evidence of accuracy using a local dataset. 

The need to substantiate the effectiveness of unsupervised machine learning algorithms for the generation of keywords of research work at the National University of the Altiplano stems from the practicality required for the implementation of these in running work environments, these algorithms used in the different computational disciplines such as data mining with text mining and artificial intelligence with machine learning, both dedicated to the generation of keywords. However, despite the existence of these methods of knowledge generation, these algorithms are not currently used; therefore, it will be critical to provide an evaluation of the most effective methods. For this purpose, we will identify, implement and compare to verify the effectiveness extraction of keywords in undergraduate research projects at the Universidad Nacional del Altiplano de Puno.

\subsection{Methods}
The methodology of this work, was based in test algorithms that could be used for the extraction of keywords that are found when analysing information stored through information systems that manage the research works, makes us have valid and reliable information with respect to the quality of it since they are regulated procedures within the Universidad Nacional del Altiplano de Puno.  The research is non-experimental, cross-sectional since data is collected in a single moment, in a certain time.  In order to describe the results based on the information collected from each of the tests performed in this study. 
\subsubsection{Population}
The research works was made in the Universidad Nacional del Altiplano de Puno, all the research works carried out by undergraduate students, these projects that are registered in PILAR which is administered by the vice chancellor for research, from which the information was extracted in order to carry out this research, which are presented in greater detail PILAR. In this study we won’t use sampling, we need to prove the algorithms with a large amount of data as possible the data was represented in (Table I). 

    \begin{table}
    \caption{Dataset Distribution}
      \centering
        \begin{tabular}{llllll}
            \toprule
            \centering
            knowledge Area & Status \\
             \cmidrule(r){2-6}
            \multicolumn{1}{c}{}& Archived & Rejected & Project & Proposal & Completed \\
            \midrule
            Biomedical          & 17       & 5        & 320     & 133      & 973       \\
            Bussiness Economics & 32       & 10       & 286     & 162      & 465       \\
            Engineering         & 60       & 24       & 847     & 368      & 1211      \\
            Social Sciences     & 34       & 20       & 891     & 291      & 1281     \\
            \bottomrule
        \end{tabular}
    \end{table}

\subsubsection{Data Source}
For the development of this research, we obtain the export of each register of the research projects of students and graduates has been prepared, which are administered in the vice chancellor for Research of the Universidad Nacional del Altiplano de Puno, taking as an initial source the data structure (Table 2), with 7430 records detailed in Table 3, these records were exported and processed individually in text files individual in text files (.txt) due to the cost of storing in memory could cause an overflow of memory.  You could see it in (Table II)

    \begin{table}
        \caption{Data Structure}
        \centering
        \begin{tabular}{cll}
        \hline
        \textbf{N°} & \textbf{Number} & \textbf{Detail}                                      \\ \hline
        01          & Guy             & \textit{Type of research work}                       \\
        02          & Code            & \textit{Labor Code based on Data}                    \\
        03          & Title           & \textit{Title of the work registered by the author.} \\
        04          & Summary         & \textit{Summary registered by the author.}           \\
        05          & Keywords        & \textit{Keywords registered by the author. {[},{]}}  \\ \hline
        \end{tabular}
    \end{table}
\subsubsection{Data Proceessing}
To develop this research work, the information detailed in (Table I), that was 7430 records of titles detailed in (Table I), whose structure is composed of titles, abstracts and keywords in the structure declared in (Table II), these data correspond to the research works of students and graduates who opted for the thesis modality to obtain their professional degree. This information that has been processed using text files, it began its treatment using regular expression methods in order to the text and exclude the special characters and stay only with the analysis of the text, to later perform the segmentation and procedures necessary to maintain the data an operable format. After that we applied the unsupervised methods to finally extract keywords. 
\begin{figure}
    \centering
    \includegraphics[scale=.6]{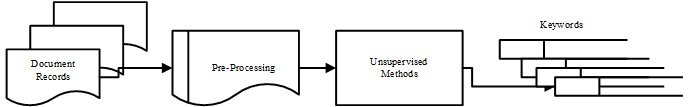}
    \caption{Data processing representation}
    \label{fig:my_label}
\end{figure}

\subsection{Model Implementation}

For the present work the models of Python Keyword Extraction PKE \cite{boudin2018a} have been used, these models were encoded with the following structure of equivalence of file names for the present research work TF-IDF(M1), KPMiner\cite{el-beltagy2010a}(M2), YAKE\cite{campos2020a}(M3), TextRank\cite{campos2020a}(M4), SingleRank\cite{wan2008a}(M5), TopicRank\cite{bougouin2013a}(M6), TopicPageRank\cite{sterckx2015a} (M7), PositionRank\cite{florescu2017a}(M8), MultipartRank\cite{boudin2018a}(M9), of these we present the evaluation and implementation in the result part. All of these models were implemented in same conditions using Python programming language.

\subsubsection{Model Assessment}

For the model assessment, we applied the following evaluation metrics (Table III) for which the sklearn library was used, as one of the most used in this type of evaluations, as well as its adaptability to the results obtained in the prediction of each group of keywords, after that we use R-Software to evaluate and get the results.

\begin{table}[]
\caption{Model Assesment}
\begin{tabular}{@{}ll@{}}
\toprule
\multicolumn{1}{c}{\textbf{Metric}} & \multicolumn{1}{c}{\textbf{Detail}}                                                                                                                                                                     \\ \midrule
F1-Score & The F1 score   can be interpreted as a harmonic mean of accuracy and recovery, where an F1   score reaches its best value at 1 and the worst score at 0.                                                \\
Recall Score & Withdrawal is   the ratio where the number of true positives and the number of false   negative. The retreat is intuitively the ability of the classifier to find   all positive samples.               \\
Precision Score & Accuracy is   the reason where the number of true positives and the number of false   positives. Accuracy is intuitively the classifier's ability not to label a   sample that is negative as positive. \\
Acuracy Score & Calculates the   accuracy of    the subassembly: The   label conjunction predicted for a sample must exactly match the corresponding   label conjunction in the original label list                     \\
Hamming Score & Hamming loss, is the number of tags that   is incorrectly predicted.                                                                                                                                    \\
Time & The time is   detailed from the reading of the source resource to the final prediction.                                                                                                                 \\
\multicolumn{2}{l}{Source: (Scikit   learn, 2021) Metrics and scoring: quantification of the quality of   predictions.  Modules.   https://scikit-learn.org/stable/modules/model\_evaluation.html}                                            \\ \bottomrule
\end{tabular}
\end{table}

\section{Results}
After implementing the methods to extract keywords, it was necessary to individually encode each one of nine models described in model implementation section in the previous part, then we have to made an analysis about time and accuracy evaluation was carried out according to the metrics detailed in (Table III) for each of the 7430 records, these prediction metrics allowed us to determine the efficiency of computed keyword extraction for each model.  In the next figure we present the main difference between these models that we proved.

\begin{figure}
    \centering
    \includegraphics[scale=.6]{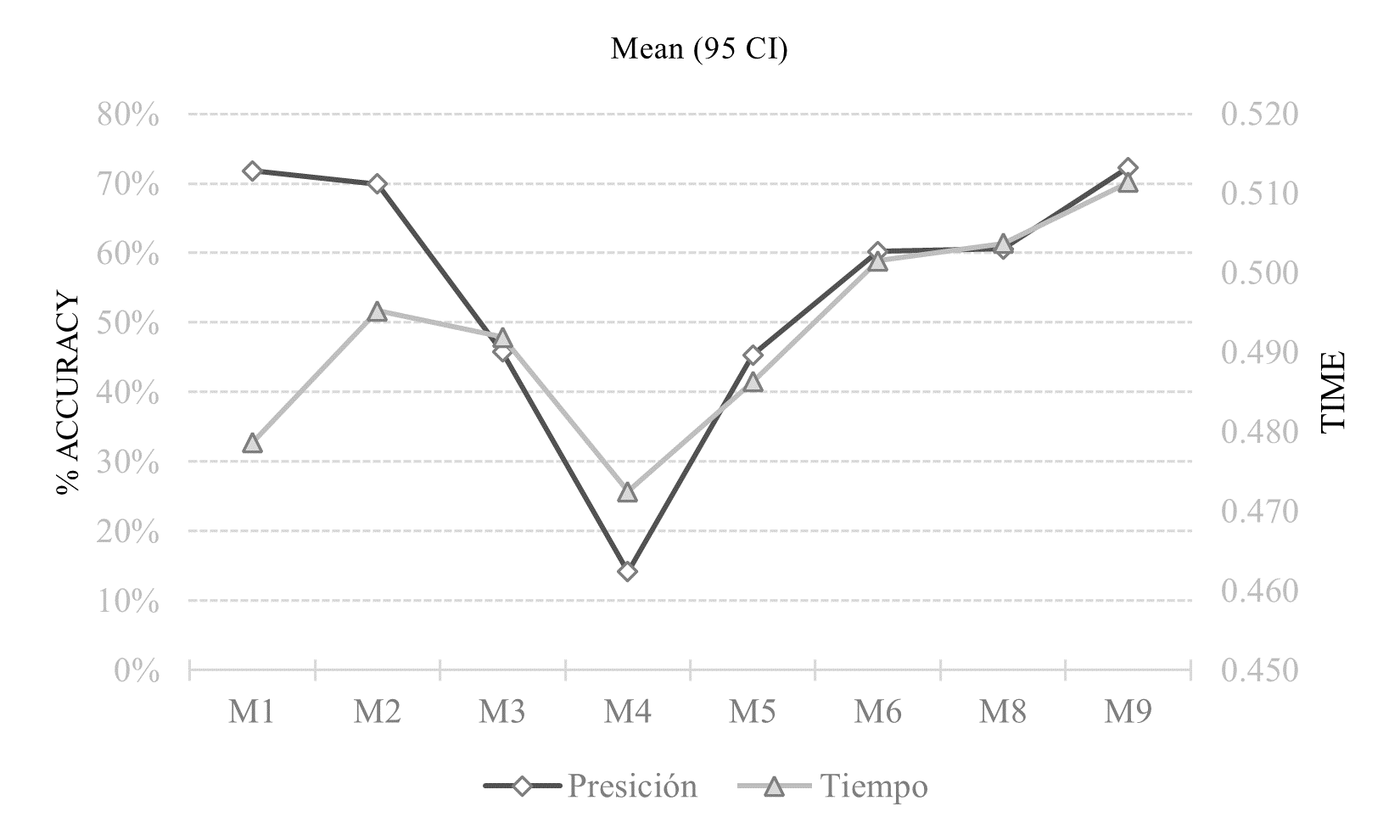}
    \caption{Models assessment by time and accuracy }
    \label{fig:my_label}
\end{figure}

As we can see at the Fig.2 with a performing a visual analysis and observing the values time and accuracy, we only choose the model M1 implemented with the algorithm TF-IDF have result with 0.4786 in time evaluation and 72\% of accuracy optimizing key word extraction work in other hand the algorithm TextRank called in this research M4 have less time 0.4725 but the accuracy of 14\% provide us evidence to discard this model in future implementations with data same as we show in this study.

\begin{table}[]
\caption{Keywords Extraction Models Acurracy}
\centering
\begin{tabular}{@{}llllll@{}}
\toprule
Model & nE   & nA   & Total & \% E  & \% A  \\ \midrule
M1    & 2094 & 5330 & 7424  & 28 \% & 72 \% \\
M2    & 2234 & 5196 & 7430  & 30 \% & 70 \% \\
M3    & 4030 & 3400 & 7430  & 54 \% & 46 \% \\
M4    & 6376 & 1054 & 7430  & 86 \% & 14 \% \\
M5    & 4063 & 3367 & 7430  & 55 \% & 45 \% \\
M6    & 2954 & 4476 & 7430  & 40 \% & 60 \% \\
M7    & 1112 & 895  & 2007  & 55 \% & 45 \% \\
M8    & 2927 & 4503 & 7430  & 39 \% & 61 \% \\
M9    & 576  & 1503 & 2079  & 28 \% & 72 \% \\ \bottomrule
\end{tabular}
\end{table}

The models based on TF-IDF, have been studied over many years always showing their efficiency, even with the improvements such as those developed in the work of Guan et al.[15], in which the recalculation of the prediction scores is carried out optimizing up to 20\% of the evaluated indicators. The same happens with the TextRank model but in a smaller range in which in several studies and especially in the one that defines its implementation its precision is demonstrated by comparing its implementation with different groups of data [19], over time several studies have been demonstrating that the implementation of this Models that are theoretically simple demonstrate their great processing capabilities. There fore we demonstrate once that TF-IDF have better result than other models, that has been implemented in this research as we can see in the last table (Table 4).

An extra challenge in this research field is use natural language processing with autoregressive language models, also build a tree with all combinations of these models to up the accuracy an reduce the time in processing always trying to get the best result in this challenge of keyword extraction.

\section{Conclusion}
The most efficient way of extracting keywords for this dataset was the TF-IDF method, due to its relationship between time and the accuracy of the analysis, because it guarantees us a shorter processing time and high precision score in keywords extraction so we can consider it a simple and general model for this and other related works.  In this research we prove nine (09) algorithms of unsupervised learning that could provide information to implementing in other universities form undergraduate thesis, also this research allowed us to exploit the computational resources in text processing, the challenge was the management and preprocessing, to finally evaluate more than fifty-six thousand records (> 56000). Finally, we show the difference between each model that we cand found for keyword extraction proposes.

\bibliographystyle{unsrt}  
\bibliography{references}

\begin{thebibliography}{10}

\bibitem{ranguelov2012a}
S.~Ranguelov.
\newblock Gestión de la información y el conocimiento en las organizaciones.
\newblock {\em Biblios}, 12(1):1–7,.

\bibitem{mart2020a}
E.~Martín-Mora, S.~Ellis, and L.M. Page.
\newblock Use of web-based species occurrence information systems by academics
  and government professionals.
\newblock {\em PLoS ONE}, 15(7).

\bibitem{tolosa2008a}
G.H. Tolosa and F.R. Bordignon.
\newblock Introducción a la recuperación de información conceptos , modelos
  y algoritmos básicos, pre-edició.
\newblock {\em Laboratorio de Redes de Datos}.

\bibitem{jayawardene2021a}
V.~Jayawardene, T.J. Huggins, R.~Prasanna, and B.~Fakhruddin.
\newblock The role of data and information quality during disaster response
  decision-making.
\newblock {\em Progress in Disaster Science}, 12:100202,.

\bibitem{torres-cruz2016a}
F.~Torres-Cruz.
\newblock Plataforma web basada en cloud computing para el seguimiento de
  proyectos de tesis de pregrado una puno 2016.

\bibitem{ahadh2021a}
A.~Ahadh, G.V. Binish, and R.~Srinivasan.
\newblock Text mining of accident reports using semi-supervised keyword
  extraction and topic modeling.
\newblock {\em Process Safety and Environmental Protection}, 155:455–465,.

\bibitem{florescu2017a}
C.~Florescu and C.~Caragea.
\newblock Positionrank: An unsupervised approach to keyphrase extraction from
  scholarly documents.
\newblock In {\em ACL 2017 - 55th Annual Meeting of the Association for
  Computational Linguistics, Proceedings of the Conference (Long Papers},
  volume~1, pages 1105–1115,.

\bibitem{xuezhong2010a}
Z.~Xuezhong, P.~Yonghong, and L.~Baoyan.
\newblock Text mining for traditional chinese medical knowledge discovery: A
  survey.
\newblock {\em Journal of Biomedical Informatics}, 43(4):650–660,.

\bibitem{xu2021a}
Z.~Xu and J.~Zhang.
\newblock Extracting keywords from texts based on word frequency and
  association features.
\newblock {\em Procedia Computer Science}, 187:77–82,.

\bibitem{gonz2012a}
M.~Gonzáles and S.~Mattar.
\newblock Las claves de las palabras clave en los artículos científicos.
\newblock {\em Revista MVZ Cordoba}, 17(2):2955–2956,.

\bibitem{mack2012a}
C.~Mack.
\newblock How to write a good scientific paper: title, abstract, and keywords.
\newblock {\em Journal of Micro/Nanolithography, MEMS, and MOEMS},
  11(2):020101,.

\bibitem{zhang2008a}
C.~Zhang, H.~Wang, Y.~Liu, D.~Wu, Y.~Liao, and B.~Wang.
\newblock Automatic keyword extraction from documents using conditional random
  fields.
\newblock {\em Journal of Computational Information Systems},
  4(3):1169–1180,.

\bibitem{kretschmann2020a}
M.~Kretschmann, A.~Fischer, and B.~Elser.
\newblock Extracting keywords from publication abstracts for an automated
  researcher recommendation system.
\newblock {\em Digitale Welt}, 4(1):20–25,.

\bibitem{campos2020a}
R.~Campos, V.~Mangaravite, A.~Pasquali, A.~Jorge, C.~Nunes, and A.~Jatowt.
\newblock Yake! keyword extraction from single documents using multiple local
  features.
\newblock {\em Information Sciences}, 509:257–289,.

\bibitem{guan2019a}
X.~Guan, Y.~Li, and H.~Gong.
\newblock Improved tf-idf for we media article keywords extraction.
\newblock {\em Journal of Physics: Conference Series}, 1302(3):032003,.

\bibitem{mahata2018a}
D.~Mahata, J.~Kuriakose, R.R. Shah, and R.~Zimmermann.
\newblock Key2vec: Automatic ranked keyphrase extraction from scientific
  articles using phrase embeddings.

\bibitem{boudin2018a}
F.~Boudin.
\newblock Unsupervised keyphrase extraction with multipartite graphs.
\newblock In {\em NAACL HLT 2018 - 2018 Conference of the North American
  Chapter of the Association for Computational Linguistics: Human Language
  Technologies - Proceedings of the Conference}, volume~2, pages 667–672,.

\bibitem{el-beltagy2010a}
S.R. El-Beltagy and A.~Rafea.
\newblock Kp-miner: Participation in semeval-2.
\newblock {\em ACL 2010 - SemEval 2010 - 5th International Workshop on Semantic
  Evaluation, Proceedings}, (July):190–193,.

\bibitem{wan2008a}
X.~Wan and J.~Xiao.
\newblock Collabrank: Towards a collaborative approach to single-document
  keyphrase extraction.
\newblock {\em Coling 2008 - 22nd International Conference on Computational
  Linguistics, Proceedings of the Conference}, 1(August):969–976,.

\bibitem{bougouin2013a}
A.~Bougouin, F.~Boudin, and B.~Daille.
\newblock Topicrank: Topic ranking for automatic keyphrase extraction.
\newblock {\em Revue Traitement Automatique des Langues}, 55(1):45–69,.

\bibitem{sterckx2015a}
L.~Sterckx, T.~Demeester, J.~Deleu, and C.~Develder.
\newblock Topical word importance for fast keyphrase extraction.
\newblock {\em WWW 2015 Companion - Proceedings of the 24th International
  Conference on World Wide Web}, (2):121–122,.

\end{thebibliography}

\end{document}